\documentclass[12pt, fleqn]{article}
\usepackage{a4wide}
\usepackage{amstext,amsfonts,amsmath,theorem}
\usepackage{graphicx}

\usepackage{verbatim}
\makeatletter \@addtoreset{equation}{section}
\@addtoreset{table}{section}

\makeatother

       {\theorembodyfont{\rmfamily}
}

       {\theorembodyfont{\rmfamily}
}

\newcommand{\G}{{\mathcal G}}
\newcommand{\ord}{\textrm{ord}}

{\end{list}}

\begin{document}
\newcounter{bean}

\thispagestyle{empty}

%\maketitle

%\vspace{-9cm}
\begin{flushright}
FTUV--06/1026 \quad IFIC/06-35 \quad Imperial/TP/2007/OV/01\\
October 26, 2006 -- March 1, 2007
\\[1cm]
\end{flushright}

%\maketitle

\begin{center}

\begin{Large}
{\bf Expansions of algebras and superalgebras \\[0.1cm] and
       some applications\footnote{Invited lecture delivered at the
       {\it Deformations and Contractions in Mathematics and Physics Workshop},
       15-21 January 2006, {\it Mathematisches Forschungsinstitut Oberwolfach} (Germany).}}
\end{Large}
\vskip 1cm

\begin{large}
J.A. de Azc\'arraga$^{a,2}$, J.M. Izquierdo$^{b,2}$, M.
Pic\'on$^{c,2}$ and O.
Varela$^{d,}$\footnote{j.a.de.azcarraga@ific.uv.es,
izquierd@fta.uva.es, picon@pd.infn.it, o.varela@imperial.ac.uk}
\end{large}
\vspace*{0.6cm}

\begin{it}
$^a$ Departamento de F\'{\i}sica Te\'orica e IFIC, Universidad de Valencia
\\
46100 Burjassot (Valencia), Spain
\\[0.4cm]
$^b$ Departamento de F\'{\i}sica Te\'orica, Universidad de Valladolid
\\
47011 Valladolid, Spain
\\[0.4cm]

$^c$ INFN Sezione di Padova and Dipartimento di Fisica
``G. Galilei'', Universit\`a di Padova
\\
Via Marzolo 8, 35131 Padova, Italy
\\[0.4cm]

$^d$ Theoretical Physics Group,
\\
The Blackett Laboratory, Imperial College, London SW7 2AZ, U.K.
\\
[0.8cm]
\end{it}
\end{center}

\begin{abstract}
After reviewing the three well-known methods to obtain Lie
algebras and superalgebras from given ones, namely, contractions,
deformations and extensions, we describe a fourth method recently
introduced, the expansion of Lie (super)algebras. Expanded
(super)algebras have, in general, larger dimensions than the
original algebra, but also include the \.In\"on\"u-Wigner and
generalized IW contractions as a particular case. As an example of
a physical application of expansions, we discuss the relation
between the possible underlying gauge symmetry of
eleven-dimensional supergravity and the superalgebra $osp(1|32)$.
\end{abstract}

\vskip 0.5cm


\vskip 0.5cm

\section{Introduction}
\label{int}

Different constructions describing the symmetry of physical theories
have made their way into physics, gradually avoiding previously
established `no-go theorems'. That was, for instance, the case of
Lie superalgebras, nowadays ubiquitous in theoretical physics and
brought into the picture as a way of mixing spacetime and internal
symmetries, not allowed in a purely bosonic context (see
\cite{Dyson}). This led to the advent of supersymmetry, a symmetry
which involves bosons and fermions simultaneously. From a
mathematical point of view, and setting aside its Lorentz part, the
superPoincar\'e algebra is a central extension of an odd (fermionic,
spinorial) abelian algebra by the spacetime translations (see
\cite{AldAazc85, CUP}) but it is not, however, the most general spacetime
superalgebra. In fact, larger supersymmetry algebras going beyond
the restrictions of the Haag-$\L$opusza\'nski-Sohnius theorem
\cite{HLS} have appeared in connection with the description of
different physical theories. For instance, the quasi-invariance
under standard supersymmetry of the Wess-Zumino (WZ) terms of the
super-$p$-brane actions results in algebras realized by the
conserved supercharges that include additional (topological) charges
\cite{AGIT89} and that are extensions of the original supersymmetry
algebra. Also -an example to be discussed in Sec. 4- an
$osp(1|32)$-related gauge formulation of $D=11$ supergravity
\cite{CJS} requires a gauge algebra that includes an additional
fermionic generator \cite{D'A+F82}. It thus makes sense to use
supersymmetry algebras beyond the standard superPoincar\'e algebra,
and many have been introduced in various contexts, leading also to a
variety of generalized, enlarged superspaces (see
\cite{vHvP,To,D'A+F82,AGIT89,BeSe95,Se97,CAIPB00,SB} and references
therein).

New algebras and superalgebras may be related to, or derived from,
previously known ones. With this in mind, we first comment on the
three well known ways to obtain new (super)algebras from given
ones, {\it i.e.}~contractions, deformations and extensions of Lie
and super Lie algebras. Then, we describe in Sec. \ref{exp} a new
procedure \cite{HS,AIPV} (see also \cite{AIPV04}), that {\it
includes} the \.In\"on\"u-Wigner (IW) and generalized
contractions, the method of Lie (super)algebra expansions, which
makes use of the geometrical structure of the algebra as expressed
by the Maurer-Cartan one-forms. At the end of Sec.~\ref{exp} the
very recent method of $S$-expansions of Lie (super)algebras
\cite{IzRoSal06} is also briefly described. We conclude with an
application in Sec. \ref{sugra}, where we show how our expanded
algebras appear \cite{BAIPV04} in the discussion of the relation
between $OSp(1|32)$ and the possible underlying gauge symmetry
group of $D=11$ supergravity \cite{CJS}.

\section{Lie algebras and superalgebras from given ones}
\label{algder}

Let $\mathcal{G}$ be a finite-dimensional Lie (super)algebra with
basis $\{X_i \}$, which may be realized by left-invariant (LI)
generators $X_i(g)$ on the corresponding (super)group manifold $G$
with local coordinates $g^i$, $i=1, \ldots ,\textrm{dim}\,G =
\textrm{dim}\, {\cal G}$. Let $c_{ij}^k$ be the structure constants
of ${\cal G}$ in the basis $\{X_i \}$, $[X_i,X_j]=c_{ij}^k X_k$. Let
$\{ \omega ^i (g) \}$, $i=1,\dots,\mathrm{dim} \, G$, be the basis
determined by the (dual, LI) Maurer-Cartan (MC) one-forms on $G$.
The MC equations that characterize ${\cal G}$, in a way dual to its
Lie bracket description, are given by
\begin{equation} \label{eq:mc}
d\omega^k(g)=-\frac{1}{2}c_{ij}^k \omega^i(g) \wedge \omega^j(g)
\; , \quad i,j,k=1,\ldots, \textrm{dim}\, {\cal G} \; .
\end{equation}

The standard procedures to obtain new (super)algebras from given
ones are:

\vskip 0.5cm \noindent {\it (a) Contractions}

Contractions go back to the work of Segal, In\"on\"u and Wigner (see
\cite{Seg51,IW53,Sal61}). In its simplest \.In\"on\"u-Wigner (IW)
form \cite{IW53}, the contraction of $\mathcal{G}$ with respect to a
subalgebra $\mathcal{L}_0 \subset {\mathcal G}$ is performed by
rescaling the generators of the coset $\mathcal{G}/\mathcal{L}_0$,
and then by taking a singular limit for the rescaling parameter.
This procedure may be extended to generalized IW contractions in the
sense of Weimar-Woods (W-W) \cite{Wei:00}. These are defined when
the vector space $W$ of $\mathcal{G}$ can be split as a sum of $n+1$
subspaces
\begin{equation}
\G \; : \; W=V_0\oplus V_1\oplus \dots\oplus V_n= \oplus_{s} V_s
\quad , \qquad s=0,1,\ldots , n \,, \label{gc1}
\end{equation}
such that the following W-W conditions are satisfied:
\begin{equation}
    c^{k_s}_{i_pj_q}=0 \;\; \text{if} \ s>p+q \qquad
\text{i.e.}\qquad  [V_p,V_q]\subset \oplus_{ s \le p+q } V_s \ ,
\quad p,q=0,1,\ldots,n \;, \label{gc2}
\end{equation}
where $i_p=1,\ldots, \text{dim} V_p$ labels the generators of
$\mathcal{G}$ in $V_p$ (we have written above $s \le p+q $ rather
than $s \le \textrm{min}\{p+q,n\}$ for simplicity.) Clearly,
condition (\ref{gc2}) implies that $V_0$ is a subalgebra ${\cal
L}_0$ of ${\cal G}$. The contracted algebra $\mathcal{G}_c$ is
obtained after the generators of each subspace are properly
re-scaled \cite{Wei:00} and a singular limit for the scaling
parameter $\lambda$ is taken. $\mathcal{G}_c$ has the same
dimension as $\mathcal{G}$; the case $n=1$ reproduces the simple
IW contraction. There have been other variations of the IW
contraction procedure (see {\it e.g.} \cite{AC79, CelTar,
Lord85,MonPat91, HMOS94}); in particular, the `graded
contractions' \cite{MonPat91} may be expressed as generalized IW
ones (see \cite{Wei:00} and the contribution of E. Weimar-Woods to
these proceedings). All contractions have in common that
$\mathcal{G}$ and $\mathcal{G}_c$ have, necessarily, the same
dimension as vector spaces.

Well known examples of contractions relevant in physics include
the Galilei algebra as an IW contraction of the Poincar\'e
algebra, the Poincar\'e algebra as a contraction of the de Sitter
algebras \cite{Nahas67}, or the characterization of the M-theory
superalgebra \cite{To} as a contraction ({\it ignoring} the
Lorentz part, {\it cf.} \cite{AIPV}) of $osp(1|32)$.

The contraction process has also been considered for `quantum'
algebras (see {\it e.g.}, \cite{CGST92}) and used, in particular, to
obtain the $\kappa$-Poincar\'e
\cite{Lukierski:1992dt} and $\kappa$-Galilei algebras
\cite{deAzcarraga:1996jp}.

\vskip 0.5cm \noindent {\it (b) Deformations}

 From a physical point of view, Lie algebra deformations
  \cite{Gerst64,NijRich66,NijRich67b,Rich67} can be
regarded as a process inverse to contractions (see also
\cite{Wei:00,Nahas67,Her70,Gil72}). Mathematically, a deformation
$\mathcal{G}_d$ of a Lie algebra $\mathcal{G}$ is a Lie algebra
`close', but not isomorphic, to $\mathcal{G}$. As in the case of
$\mathcal{G}_c$ above, $\mathcal{G}_d$ has the same dimension as
$\mathcal{G}$.

Deformations are obtained by modifying the {\it r.h.s.} of the
original commutators by adding new terms that depend on a
parameter $t$ in the form
\begin{equation}
 [X,Y]_t=[X,Y]_0+\sum^\infty_{i=1} \omega_i(X,Y) t^i \ , \quad
X,Y \in {\mathcal G}\;,\quad \omega_i(X,Y) \in {\mathcal G} \; .
\label{def1}
\end{equation}
Checking the Jacobi identities up to $O(t^2)$, it is seen that the
expression satisfied by $\omega_1$ characterizes it as a
two-cocycle. Thus, the second Lie algebra cohomology group
$H^2(\mathcal{G}, \mathcal{G})$ of $\mathcal{G}$ with coefficients
in the Lie algebra $\mathcal{G}$ itself is the group of
infinitesimal deformations of $\mathcal{G}$ and $H^2(\mathcal{G},
\mathcal{G})=0$ is a {\it sufficient} condition for rigidity
\cite{Gerst64,NijRich66, Rich67}. In this case, $\mathcal{G}$ is
{\it rigid} or {\it stable} under infinitesimal deformations; any
attempt to deform it yields an isomorphic algebra. The problem of
finite deformations depends on the integrability of the
infinitesimal deformation; the obstruction is governed by
$H^3(\mathcal{G}, \mathcal{G})$ which needs being trivial.

As is well known, the Poincar\'e algebra may be seen as  a
deformation of the Galilei one, a fact that may be viewed as a
group theoretical prediction of relativity. The de Sitter,
$so(4,1)$, and anti de Sitter, $so(3,2)$, algebras are
stabilizations of the Poincar\'e algebra; $osp(1|4)$ is a
deformation of the $N=1$, $D=4$ superPoincar\'e algebra
\cite{B86}. Quantization itself may also be looked at as a
deformation (see \cite{Mo49,FLS76,Vey75}), the classical limit
being the contraction limit $\hbar \rightarrow 0$. Nontrivial
central extensions of Lie algebras may also be considered as
deformations or partial stabilizations of trivial (direct sum)
extensions.

\vskip 0.5cm
 \noindent{\it (c) Extensions (of a Lie algebra or
superalgebra by another one)}

 In contrast with the  procedures (a) and (b) above, the initial data
of the extension problem includes {\it two} Lie algebras
${\mathcal G}$ and ${\mathcal A}$. A Lie algebra
$\tilde{\mathcal{G}}$ is an extension of $\mathcal{G}$ by
$\mathcal{A}$ if $\mathcal{A}$ is an ideal of
$\tilde{\mathcal{G}}$ and
$\tilde{\mathcal{G}}/\mathcal{A}=\mathcal{G}$. As a result,
$\text{dim}\, \tilde{\mathcal{G}}= \text{dim}\, \mathcal{G} +
\text{dim}\, \mathcal{A}$, so that the extension process is also
`dimension preserving'. To obtain an extension
$\tilde{\mathcal{G}}$ of $\mathcal{G}$ by $\mathcal{A}$ it is
necessary to specify first an action $\rho$ of $\mathcal{G}$ on
$\mathcal{A}$ {\it i.e.}, a Lie algebra homomorphism $\rho: \,
\mathcal{G} \longrightarrow \text{End}\, \mathcal{A}$. The
possible extensions $\tilde{\mathcal{G}}$ for a given set
$(\mathcal{G},\mathcal{A} ,\rho)$ and the possible obstructions to
the extension process are, again, governed by cohomology (see
\cite{CUP} for full details and references).

Examples of extensions in physics are the centrally extended
Galilei algebra, which is relevant in non-relativistic quantum
mechanics (and that may be obtanined as a contraction of the
trivially extended $D=4$ Poincar\'e group, see \cite{AA85} to see
how contractions may generate cohomology), the two-dimensional
extended Poincar\'e algebra that allows \cite{CJ92} for a gauge
theoretical derivation of the Callan-Giddings-Harvey-Strominger
model \cite{Callan:1992rs} for two-dimensional gravity, or the
M-theory superalgebra that, without its Lorentz automorphisms
part, is the maximal central extension of the abelian $D=11$
supertranslations algebra (\cite{vHvP,To,AGIT89,CAIPB00}).

\medskip
We now turn to a new procedure, the expansion of Lie algebras and
superalgebras.

\section{Expansions of Lie (super)algebras}
\label{exp}

Under a different name, Lie algebra expansions were first used in
\cite{HS}, and then the method was studied in general in
\cite{AIPV} (see also \cite{AIPV04}). The idea is to perform a
rescaling by a parameter $\lambda$ of some of the group
coordinates $g^i$, $i=1,\dots,\text{dim}\, \mathcal{G}$.
Consequently, the MC one-forms $\omega^i(g,\lambda)$ of
$\mathcal{G}$ are expanded as power series in $\lambda$. Inserting
these expansions (polynomials in $\lambda$) in the original MC
equations for $\mathcal{G}$, one obtains a set of equations that
have to be satisfied, each one corresponding to a power of
$\lambda$. The problem at this stage is how to cut the series
expansions of the different $\omega^i$'s in such a way that the
resulting set of MC-like equations be closed under $d$, so that it
defines the MC equations of a new, finite-dimensional {\it
expanded} Lie algebra.

In fact, notice that it is possible to write the MC forms
$\omega^i(g)$ of $\cal G$ as polynomials in the group coordinates
$g^i$ (see \cite{AIPV}) as
\begin{equation} \label{eq:serie2}
\omega^i(g) =  \left[ \delta_j^i +\frac{1}{2!} c_{j k}^i g^k+
   \sum_{n=2}^{\infty}
\frac{1}{(n+1)!} c_{j k_1}^{h_1}c_{h_1 k_2}^{h_2} \ldots
       c_{h_{n-2} k_{n-1}}^{h_{n-1}}c_{h_{n-1} k_n}^{i}
g^{k_1} g^{k_2} \ldots g^{k_{n-1}} g^{k_n} \right] dg^j \ .
\end{equation}
Hence, a redefinition
\begin{equation}
g^l \rightarrow \lambda^{q{}_l} g^l
\end{equation}
of {\it some} group coordinates $g^l$ will produce an expansion of
the MC one-forms $\omega^i(g, \lambda)$ as a sum of one-forms
$\omega^{i,\alpha}(g)$ on $G$ multiplied by the corresponding
powers $\lambda^\alpha$ of $\lambda$. The actual form of the power
series of $\omega^i(g,\lambda)$ is in fact dependent on the
possible structure of $\cal G$ if a suitable redefinition of the
group parameters is made. In general, moreover, the richer the
structure of $\cal G$, the more possibilities arise to cut the
$\omega^i$'s power series in order to obtain well defined
finite-dimensional Lie algebras.

For the sake of definiteness, let us discuss the case in which
$\cal G$ satisfies the Weimar-Woods (W-W) conditions (\ref{gc1}),
(\ref{gc2}), referring to \cite{AIPV} for other interesting cases.
When the W-W conditions are satisfied, the MC one-forms of $\cal
G$ arrange themselves in $n+1$ sets $\{\omega^{i_p}\}$,
$i_p=1,\ldots, \text{dim} V_p$, $p=0,1, \ldots , n$, corresponding
to each subspace $V_p$ in (\ref{gc1}), and the structure constants
of ${\cal G}$ satisfy $c^{k_s}_{i_pj_q}=0$ if $s>p+q$; the
subspace $V_0$ is a subalgebra ${\cal L}_0$ of ${\cal G}$.
Consider next the rescaling $g^{i_{ p}} \rightarrow \lambda^p
g^{i_{ p}}$, $p=0,\ldots,n$, or, explicitly,
\begin{eqnarray} \label{eq:nredef} && g^{i_{ 0}} \rightarrow
g^{i_{ 0}} \; , \quad  g^{i_{ 1}} \rightarrow \lambda g^{i_{ 1}}
\; , \; \ldots \; , \quad  g^{i_{ n}} \rightarrow \lambda^n g^{i_{
n}} \; ,
\end{eqnarray}
of the group parameters, where $g^{i_{ p}}$ is subordinated to the
splitting (\ref{gc1}) in an obvious way. With this rescaling, the
condition (\ref{gc2}), namely, $ c^{k_s}_{i_pj_q}=0$ if $s>p+q$,
produces that the series expansion of the forms $\omega^{i_p}$ in
each subspace $V_p$ that results from the insertion of
(\ref{eq:nredef}) in (\ref{eq:serie2}), starts with the power
$\lambda^p$, $p=0,1, \ldots , n$ \cite{AIPV}:
\begin{equation}
\omega^{i_p}=\sum^\infty_{\alpha_p=p}
\omega^{i_p,\alpha_p}\lambda^{\alpha_p} = \lambda^p \omega^{i_p ,
p} + \lambda^{p+1} \omega^{i_p , p+1} + \ldots \label{expWW} \; ,
\end{equation}
where the index denoting each power of $\lambda$ has been written
as $\alpha_p$ to stress the fact that the series expansion is
different for each $\omega^{i_p}$, $p=0,1, \ldots , n$.

Inserting the series (\ref{expWW}) into the MC equations
(\ref{eq:mc}) of ${\cal G}$ and equating the coefficients with the
same powers of $\lambda$, a set of equations for the various
coefficient one-forms $\omega^{i_p , \alpha_p}$ is obtained:
\begin{equation} \label{eq:MCn}
d\omega^{k_{ s}, \alpha_s}= -\frac{1}{2} C_{i_{ p},\beta_p \;
j_{q},\gamma_q}^{k_{ s},\alpha_s}\; \omega^{i_{ p}, \beta_p}
\wedge \omega^{j_{ q}, \gamma_q} \quad ,
\end{equation}
where
\begin{equation} \label{eq:Cn}
C_{i_{ p},\beta_p \; j_{ q},\gamma_q}^{k_{ s},\alpha_s}= \left\{
\begin{array}{lll} 0, &
\mathrm{if} \ \beta_p + \gamma_q \neq \alpha_s  \\
c_{i_{ p}j_{ q}}^{k_{s}}, & \mathrm{if} \ \beta_p + \gamma_q =
\alpha_s  \end{array} \right. \quad \begin{array}{l}
p,q,s=0,1,\ldots, n \\
i_{p,q,s}=1,2,\ldots, \textrm{dim} \, V_{p,q,s} \\
\alpha_p,\beta_p,\gamma_p=p,p+1, \ldots, N_p
\end{array}
\end{equation}
and the $c_{i_p j_q}^{k_s}$ satisfy (\ref{gc2}).

One may now consider whether the series (\ref{expWW}) for each
$\omega^{i_p}$ may be cut at an arbitrary order $N_p$ {\it i.e.},
whether any finite number of one-form coefficients $\omega^{i_p ,
\alpha_p}$, $\alpha_p = p , p+1 , \ldots N_p$, can be retained in
such a way that equations (\ref{eq:MCn}), (\ref{eq:Cn}) define,
respectively, the MC equations and structure constants of a new,
finite-dimensional Lie algebra labelled ${\cal G}(N_0, \ldots ,
N_n)$. This is clearly not the case. For ${\cal G}(N_0, \ldots ,
N_n)$ to be a Lie algebra, two conditions must be met:

\vskip 10pt

\noindent a) the set of retained one-form coefficients,
\begin{equation}
\label{newomegas} \{ \omega^{i_{ 0}, 0}, \omega^{i_{ 0}, 1},
\stackrel{N_0 +1}{\ldots}, \omega^{i_{ 0}, N_{ 0}};\, \,
\omega^{i_{ 1}, 1}, \stackrel{N_1}{\ldots}, \omega^{i_{ 1}, N_{
1}}; \,\, \ldots; \,\, \omega^{i_{n}, n},
\stackrel{N_n-n+1}{\ldots}, \omega^{i_{n}, N_{n}} \} \; ,
\end{equation}
which determines the dimension of the expanded algebra ${\cal
G}(N_0, \ldots , N_n)$ by
\begin{equation} \label{eq:dim}
\textrm{dim} \, \mathcal{G}(N_0, \ldots, N_n) = \sum_{p=0}^{n}
(N_p -p+1) \, \textrm{dim} \, V_p \;,
\end{equation}
must be closed under the exterior differential $d$; and

\vskip 10pt

\noindent b) the symbols $C_{i_{ p},\beta_p \; j_{
q},\gamma_q}^{k_{ s},\alpha_s}$ defined in (\ref{eq:Cn}) must obey
the Jacobi identity (notice that their definition (\ref{eq:Cn})
makes them already inherit the symmetry properties of the
structure constants $c_{i_p j_q}^{k_s}$ of the original
(super)algebra).

\vskip 10pt

With regard to the condition a) notice that, due to the W-W
conditions (\ref{gc2}), the forms $\omega^{i_p , \beta_p}$ that
enter the expression of $d\omega^{k_s , \alpha_s}$ in (\ref{eq:MCn})
are those with \cite{AIPV}
\begin{equation} \label{betal}
\beta_p \leq  \left\{
\begin{array}{lll} \alpha_s -s + p, &
\mathrm{if} \ p \leq s   \\
\alpha_s, & \mathrm{if} \ p > s
\end{array} \right. \quad \begin{array}{l}
p,s=0,1,\ldots, n \\
\alpha_p,\beta_p=p,p+1, \ldots, N_p
\end{array} \; .
\end{equation}
Hence, the set of forms (\ref{newomegas}) will be closed under $d$
if the cutting orders satisfy
\begin{equation} \label{NpNs}
N_p \geq \left\{
\begin{array}{lll} N_s -s + p, &
\mathrm{if} \ p \leq s   \\
N_s, & \mathrm{if} \ p > s
\end{array} \right. \quad \begin{array}{l}
p,s=0,1,\ldots, n \\
\alpha_p,\beta_p=p,p+1, \ldots, N_p
\end{array} \; ,
\end{equation}
namely, when
\begin{equation}
N_{p+1}=N_p \qquad \text{or} \qquad N_{p+1}=N_p +1 \qquad
(p=0,1,\ldots, n-1)\ \ ,  \label{conWW}
\end{equation}
which gives \cite{AIPV} $2^n$  possibilities in all.

 As for the condition b), the Jacobi identities
{\setlength\arraycolsep{2pt}
\begin{eqnarray}\label{eq:granjacobi}
&& C_{i_p,\beta_p\;[j_q,\gamma_q}^{k_s,\alpha_s} C_{l_t,\rho_t\;
m_u,\sigma_u]}^{i_p,\beta_p} =0 = \nonumber \\
&&  C_{i_p,\beta_p\;j_q,\gamma_q}^{k_s,\alpha_s} C_{l_t,\rho_t\;
m_u,\sigma_u}^{i_p,\beta_p}  +
C_{i_p,\beta_p\;m_u,\sigma_u}^{k_s,\alpha_s}
C_{j_q,\gamma_q\;l_t,\rho_t}^{i_p,\beta_p}+
C_{i_p,\beta_p\;l_t,\rho_t}^{k_s,\alpha_s}
C_{m_u,\sigma_u\;j_q,\gamma_q}^{i_p,\beta_p} \; ,
\end{eqnarray}
}are satisfied through those for $\mathcal{G}$. This is a
consequence of the fact that, for ${\mathcal G}$, the exterior
derivative of the $\lambda$-expansion of the MC equations is the
$\lambda$-expansion of their exterior derivative, but it may also
be seen directly.

Indeed, we only need to check that (\ref{eq:granjacobi}) reduces
to the Jacobi identities for $\mathcal{G}$ when the order in the
upper index is the sum of those in the lower ones since the $C$'s
are zero otherwise. First we see that, when
$\alpha_s=\gamma_q+\rho_t+\sigma_u$, all three terms in the
r.h.s.~of (\ref{eq:granjacobi}) give non-zero contributions. This
is so because the range of $\beta_p$ is only limited by
$\beta_p\leq\alpha_s$, which holds when $\beta_p=\rho_t+\sigma_u$,
$\beta_p=\gamma_q+\rho_t$ and $\beta_p=\sigma_u+\gamma_q$.
Secondly, and since $\beta_p\geq p$, we also need that the terms
in the $i_p$ sum that are suppressed in (\ref{eq:granjacobi}) when
$p>\beta_p$ be also absent in the Jacobi identities for
$\mathcal{G}$ so that (\ref{eq:granjacobi}) does reduce to the
Jacobi identities for $\mathcal{G}$. Consider {\it e.g.}, the
first term in the r.h.s.~of (\ref{eq:granjacobi}). If $p>\beta_p$,
then $p>\rho_t+\sigma_u$ and hence $p>t+u$. Thus, by the W-W
condition (\ref{gc2}), this term will not contribute to the Jacobi
identities for $\mathcal{G}$ and no sum over the subspace $V_p$
index $i_p$ will be lost as a result. The argument also applies to
the other two terms for their corresponding $\beta_p$'s.

A particular solution to (\ref{conWW}) is obtained by setting $N_p
= p$, $p=0,1,\ldots,n$, which defines $\mathcal{G}(0,1,\ldots,
n)$, with $\textrm{dim} \,\mathcal{G}(0,1,\ldots,
n)=\textrm{dim}\,\mathcal{G}$ by (\ref{eq:dim}). Since in this
case $\alpha_p$ takes only one value ($\alpha_p=N_p=p$) for each
$p=0,1,\ldots,n$, we may drop this label. Then, the structure
constants (\ref{eq:Cn}) for $\mathcal{G}(0,1, \ldots,n)$ read
\begin{equation} \label{eq:CIWn}
C_{i_{ p}\, j_{ q}}^{k_{ s}}= \left\{ \begin{array}{lll} 0, &
\mathrm{if} \ p + q \neq s  \\
c_{i_{ p}j_{ q}}^{k_{s}}, & \mathrm{if} \ p + q = s  \end{array}
\right. \quad \begin{array}{l}
p=0,1,\ldots,n \\
i_{p,q,s}=1,2,\ldots, \textrm{dim} \, V_{p,q,s} \; ,
\end{array}
\end{equation}
which shows that $\mathcal{G}(0,1,\ldots,n)$ is the generalized IW
contraction of $\mathcal{G}$, in the sense of \cite{Wei:00},
subordinated to the splitting (\ref{gc1}). Obviously, if $n=1$,
${\cal G} = {\cal L}_0 \oplus V_1$, where ${\cal L}_0$ is a
subalgebra, and the simple IW contraction is recovered as the
expansion ${\cal G} (0,1)$.

Thus, we have actually proved the following

\vspace{0.5cm}

{\bf Theorem 1.} Let $\mathcal{G}=V_{0} \oplus V_{1} \oplus \cdots
\oplus V_n$ be a splitting of $\mathcal{G}$ into $n+1$ subspaces.
Let $\mathcal{G}$ fulfil the Weimar-Woods contraction condition
(\ref{gc2}) subordinated to this splitting, $c_{i_p j_q}^{k_s}=0$
if $s>p+q$. The one-form coefficients $\omega^{i_p , \alpha_p}$ of
(\ref{newomegas}) resulting from the expansion of the
Maurer-Cartan forms $\omega^{i_p}$ in which $g^{i_{ p}}
\rightarrow \lambda^p g^{i_{ p}}, \; p=0,\ldots,n$
(eq.~(\ref{eq:nredef})), determine expanded Lie algebras, denoted
$\mathcal{G}(N_0,N_1,\ldots,N_n)$, of dimension (\ref{eq:dim}) and
structure constants given by
\begin{equation}\label{stconsWW}
C_{i_{ p},\beta_p \; j_{ q},\gamma_q}^{k_{ s},\alpha_s}= \left\{
\begin{array}{lll} 0, &
\mathrm{if} \ \beta_p + \gamma_q \neq \alpha_s  \\
c_{i_{ p}j_{ q}}^{k_{s}}, & \mathrm{if} \ \beta_p + \gamma_q =
\alpha_s  \end{array} \right. \quad \begin{array}{l}
p,q,s=0,1,\ldots, n \\
i_{p,q,s}=1,2,\ldots, \textrm{dim} \, V_{p,q,s} \\
\alpha_p,\beta_p,\gamma_p=p,p+1, \ldots, N_p \quad ,
\end{array}
\end{equation}
(eq. (\ref{eq:Cn})) if $N_p=N_{p+1}$ {\it or} $N_p=N_{p+1}-1$
($p=0,1,\ldots,n-1$) in $(N_0,N_1,\ldots,N_n)$. In particular, the
$N_p=p$ solution determines the algebra $\mathcal{G}(0,1,\ldots,
n)$, which is the generalized \.In\"on\"u-Wigner contraction of
$\mathcal{G}$.

\vspace{0.5cm}

In general, the Lie algebra $\mathcal{G}(N_0,N_1,\ldots,N_n)$ is
larger than ${\cal G}$ (see equation (\ref{eq:dim})). This fact,
and its derivation, justifies the name of {\it expanded} algebras
\cite{AIPV}.

An interesting case is that of Lie superalgebras, the splitting of
which into subspaces naturally satisfies the W-W conditions. For
instance, we may take $\mathcal{G}=V_0\oplus V_1$ or
$\mathcal{G}=V_0 \oplus V_1 \oplus V_2$ with $V_0$ or $V_0\oplus
V_2$ containing all the even (bosonic) generators and $V_1$
containing the Grassmann odd (fermionic) ones. Then, the
expansions of the MC one-forms of $V_1$ ($V_0$ and $V_2$) only
contain odd (even) powers of $\lambda$ \cite{AIPV}. The
consistency conditions for the existence of
$\mathcal{G}(N_0,N_1)$-type expanded superalgebras require that
\begin{equation}
    N_0=N_1-1\ \quad \textrm{or} \quad \ N_0=N_1+1 \quad ,
\label{consup1}
\end{equation}
and, for the $\mathcal{G}(N_0,N_1,N_2)$ case, that one of the
three following possibilities holds:
\begin{equation}
N_0=N_1+1=N_2\ , \ \; N_0=N_1-1=N_2\ , \ \; N_0=N_1-1=N_2- 2 \quad
. \label{consup2}
\end{equation}
This last case allows us to obtain, for example, the M-algebra
including the Lorentz $SO(1,10)$ automorphisms as the expansion
$osp(1|32)(2,1,2)$ of $osp(1|32)$. The appropriate choice of
$V_0$, $V_1$, $V_2$ leading to this expansion can be found in
\cite{AIPV}.\\

{\bf $S$-expansions of Lie (super)algebras}

As we have seen, the expansion method allows us to obtain new Lie
algebras of increasing dimensions from $\cal G$ by a geometric
procedure based on expanding the MC forms. One may think of other
possibilities leading, in general, to larger algebras from a given
one. We conclude this section by briefly describing another
construction, very recently proposed \cite{IzRoSal06}, which is
based on combining the structure constants of $\G$ with the inner
law of a semigroup $S$ to define the Lie bracket of a new,
$S$-expanded algebra. The ingredients here are, then, the algebra
$\cal G$ and a certain semigroup $S$.

Consider a finite abelian semigroup $S$ (a set $S$ with $\ord \,S$
elements $\alpha, \beta, \gamma, \ldots\in S$, endowed with a
commutative and associative composition law $S \times S
\rightarrow S$, $(\alpha , \beta) \mapsto \alpha \beta = \beta
\alpha$). Then, one may define a Lie algebra structure over the
vector space obtained by taking $\ord \,S$ copies of $\G$,
\begin{equation}
\label{dimSexp} \G_S: W_\alpha \oplus W_\beta \oplus W_\gamma
\oplus \cdots =\oplus_{\alpha \in S} W_\alpha \quad
 (W_\alpha \approx\G\quad \forall \alpha), \quad
 \textrm{dim} \ {\cal G}_S =
\textrm{ord} \ S \times \textrm{dim} \ {\cal G}  \; ,
\end{equation}
by means of the structure constants
\begin{equation} \label{scdirsum1}
C_{i \alpha \ j \beta}^{k \gamma}= c_{ij}^k \,\delta_{\alpha
\beta}^\gamma  \; ,
\end{equation}
where $\delta$ is the Kronecker symbol and the subindex $\alpha
\beta \in S$ denotes the inner composition in $S$ so that
$\delta_{\alpha\beta}^\gamma=1$ when $\alpha\beta=\gamma$ in $S$ and
zero otherwise. The constants $C_{i \alpha \ j \beta}^{k \gamma}$
defined by (\ref{scdirsum1}) inherit the symmetry properties of the
$c_{ij}^k$ of ${\cal G}$ by virtue of the abelian character of the
$S$-product, and satisfy the Jacobi identity $C_{[i \alpha \ j
\beta}^{h \delta} C_{ k \gamma ] \ h \delta }^{l \epsilon} =0$
because of the commutativity and associativity of the semigroup
inner law and the Jacobi identity of $\cal G$, $c_{[i j }^{h } c_{ k
] h}^{l} =0$. This Lie (super)algebra ${\cal G}_S$ was called {\it
$S$-expansion} of $\cal G$ \cite{IzRoSal06}.

When the Lie brackets of the original algebra $\G$ satisfy certain
conditions, as {\it e.g.} the W-W conditions (\ref{gc2}), then
certain subalgebras $\G_S^\prime$ can be extracted
\cite{IzRoSal06} from the $S$-expanded algebra $\G_S$ provided
that it is possible to find subsets of $S$ (see (\ref{Ssplit})
below) the composition of which {\it mimics} the subspace
structure of $\G$ with respect to its Lie bracket (see eq.
(\ref{gc2})). These $\G_S^\prime$ can then be used to retrieve the
expansions ${\cal G}(N_0,\ldots,N_n)$. The procedure is not
entirely straightforward, so we shall make explicit the
intermediate steps below.

Let then $\G$ satisfy the W-W conditions (\ref{gc2}) and let us
conveniently choose the semigroup $S$ as \cite{IzRoSal06}
\begin{equation}
\label{SN1mult} S=\{\alpha \ | \ \alpha = 0, 1, \ldots , N, N+1
\}\;,\qquad \alpha \beta = \left\{
\begin{array}{lll}
\alpha + \beta, & \mathrm{if} \ \alpha + \beta < N+1  \\
N+1, & \mathrm{if} \ \alpha + \beta \geq N+1
\end{array} \right. \; ,
\end{equation}
where $\alpha+\beta$ is simply the sum of natural numbers. The
underlying vector space of any $S$-expanded algebra is
$\G_S=
\oplus_{\alpha \in S} W_\alpha\,$; each copy $W_\alpha$ of the vector
space $W$ of ${\cal G}$ obviously admits the same splitting,
$W_\alpha = \oplus_{p} V_{p\alpha}, \; p=0,\ldots,n$. Hence, the
$\G_S$ vector subspace structure splits as
\begin{equation}
\label{splitSexp} \G_S \; = \oplus_{p} \oplus_{\alpha \in S} V_{p
\alpha} \; , \quad \textrm{where} \;\; V_{p \alpha} \approx V_p
\qquad p=0,1,\ldots,n\; ,\qquad \alpha \in S \; .
\end{equation}
As ${\cal G}$ satisfies the W-W conditions,
$[V_p , V_q ] \subset \oplus_{ s \le p+q } V_s \, , \;
p,q=0,1,\ldots n \; ,$ the Lie bracket subspace structure of $\G_S$
inherited from that of ${\cal G}$ is
\begin{equation} \label{bracksplitSexp}
\G_S \; : \; [V_{p \alpha} , V_{q \beta} ] \subset \oplus_{ s \le
p+q} V_{s  \alpha \beta} \; ,
\end{equation}
where $\alpha \beta$ in $V_{s  \alpha \beta}$ again denotes
$S$-composition (here again, and also below, we write
$s \le p+q $ rather than $s \le \textrm{min}\{p+q,n\}$ for simplicity's
sake).

Let $\{S_s\}$ in
\begin{equation} \label{Ssplit}
S= \cup_{s} S_s \;, \quad s=0,1,\ldots,n\;,
\end{equation}
be a (not necessarily disjoint) collection of subsets $S_s\subset
S$ (compare (\ref{Ssplit}) and (\ref{gc1})). The subsets
$S_s\subset S$ are thus in one-to-one correspondence with the
vector subspaces $V_s \subset \G$ in (\ref{gc1}). When the
condition
\begin{equation}
\label{multsplitSexp} S_p  S_q \subset \cap_{ s \le p+q} S_s \quad
, \qquad S_p  S_q := \{ \alpha_p \beta_q \ | \ \alpha_p \in S_p ,
\,\beta_q \in S_q \} \; ,
\end{equation}
is satisfied, the collection of subsets $S_s\subset S$ is adapted
to the partition $V_s\subset \G$ of the Lie algebra in the sense
that eqs. (\ref{gc1}) and (\ref{Ssplit}) induce similar structures
in eqs. (\ref{gc2}) and (\ref{multsplitSexp}) respectively. Such a
collection $\{S_s\}\,$, $S= \cup_s S_s\,$, was said in
\cite{IzRoSal06} to be resonant with the algebra decomposition
$\G=\oplus_s V_s \, $; eq. (\ref{multsplitSexp}) was called the
resonance condition.

Now, the vector {\it sub}space of (\ref{splitSexp}) $\G_S^\prime
\subset \G_S$, defined by
\begin{equation}\label{vsgps}
\G_S^\prime \; = \; \oplus_{p} \left( \oplus_{\alpha_p \in S_p}
V_{p \alpha_p} \right)\;,\qquad p=0, \ldots, n\quad ,
\end{equation}
({\it cf}.~(\ref{splitSexp})) is actually a {\it subalgebra}
(called resonant in \cite{IzRoSal06}) of ${\cal G}_S$, $\; {\cal
G}_S^\prime \subset\G_S$, with Lie bracket structure given by
\begin{equation}
\label{stgsprime} \G_S^\prime \; : \; [V_{p \alpha_p} , V_{q
\beta_q} ] \subset \oplus_{ s \le p+q} V_{s\,  \alpha_p \beta_q}
\; , \;
\end{equation}
and with structure constants determined by (\ref{scdirsum1}) and
the $S$ inner law in eq. (\ref{SN1mult}). This is so because the
subspace structure (\ref{stgsprime}) comes from (\ref{gc2}) and
follows from (\ref{bracksplitSexp}), and the {\it r.h.s.} of
(\ref{stgsprime}) is in $\G_S^\prime$ because $\alpha_p \beta_q
\in S_s\;\; \forall s \le p+q$ due to the resonant condition
(\ref{multsplitSexp}).

 We now move on to show how the expansions ${\cal G}(N_0,\ldots,N_n)$ in
 Theorem 1 can be retrieved from the above subalgebra $\G_S^\prime$ of $\G_S$.
Let us take the following collection of subsets of $S$
\begin{equation}\label{Spartition}
S_p=\{\alpha_p \ | \ \alpha_p=p,\ldots,N+1\} \; ,\quad p=0,\dots
n\;,
\end{equation}
which clearly satisfy (\ref{multsplitSexp}). Let us split them as
$S_p={\check S}_p \cup {\hat S}_p$,
$\;{\check S}_p=\{p,\ldots,N_p\},\; {\hat S}_p=\{{N_p+1},
\ldots , {N+1}\}$ \cite{IzRoSal06} and use ${\hat S}_p$ to
introduce the vector subspace ${\hat\G}_S^\prime \subset
\G_S^\prime$ by
\begin{equation}
{\hat \G}_S^\prime \; = \; \oplus_{p} \left( \oplus_{\alpha_p \in
{\hat S}_p} V_{p \alpha_p} \right), \qquad p=0,\ldots,n
\end{equation}
({\it c.f.} (\ref{vsgps})).
%with dimension
%\begin{equation} \textrm{dim} \ {\hat \G}_S^\prime =
%\sum_{p=0}^n \sum_{\alpha_p \in {\hat S}_p} \textrm{dim} \ V_{p
%\alpha_p} = \sum_{p=0}^n \sum_{\alpha_p =N_p+1 }^{N+1} \textrm{dim}
%\ V_{p \alpha_p} = \sum_{p=0}^n (N -N_p +1) \ \textrm{dim} \ V_p \;
%.
%\end{equation}
Now, if the integers $N_p$, $p=0, \ldots ,n$, are chosen to obey
the restrictions (\ref{conWW}) or, equivalently, (\ref{NpNs}),
then ${\hat \G}_S^\prime$ is an ideal of $\G_S^\prime$. Indeed, we
see from eq. (\ref{scdirsum1}) that ${\hat \G}_S^\prime$ will be
an ideal of $\G_S^\prime$ if for $\alpha_p \in {\hat S}_p$ and
$\beta_q \in S_q$ in (\ref{stgsprime}),  $\alpha_p \beta_q \in
{\hat S}_s$ where $s= \min \{ n, p+q \}$. This is indeed the case:
if $s=p+q\leq n$ in (\ref{stgsprime}), eq. (\ref{NpNs}) for $p
\leq s$ leads to $(N_p + 1) + q \geq (N_{p+q} + 1)$, and hence
$\alpha_p \beta_q \in {\hat S}_{p+q}$. And, if $p+q > n$, $s=n$ in
(\ref{stgsprime}),  eq. (\ref{NpNs}) for $p \leq n$ gives $N_p
\geq N_n -n +p$ and so $N_p +q \geq N_n -n +p +q$. Since now $p+q
> n$, this gives $(N_p +1) +q \geq (N_n +1)$, and thus $\alpha_p
\beta_q \in {\hat S}_n$ $\forall$ $\alpha_p \in  {\hat S}_p$,
$\forall$  $\beta_q \in S_q$.

The quotient of ${\G}_S^\prime$ by the ideal ${\hat \G}_S^\prime$ ,
\begin{equation}
{\check \G}_S^\prime = {\cal G}_S^\prime / {\hat \G}_S^\prime \;
\end{equation}
defines the algebra ${\check \G}_S^\prime$ (not a subalgebra of
$\G_S^\prime$), with underlying vector space
\begin{equation}
{\check \G}_S^\prime \; = \; \oplus_{p} \left( \oplus_{\alpha_p \in
{\check S}_p} V_{p \alpha_p} \right), \qquad p=0,\ldots,n \quad .
\end{equation}
As vector spaces, ${\hat \G}_S^\prime$ and ${\check\G}_S^\prime$ are
complementary in ${\G}_S^\prime$. The dimension of ${\check
\G}_S^\prime$ is given by
\begin{eqnarray}
\label{dimcheck}
\textrm{dim} \ {\check \G}_S^\prime &=& \textrm{dim} \ {\cal
G}_S^\prime - \textrm{dim} \ {\hat \G}_S^\prime \nonumber  \\
&=& \sum_{p=0}^n \sum_{\alpha_p \in {\check S}_p} \textrm{dim} \
V_{p \alpha_p} = \sum_{p=0}^n \sum_{\alpha_p =p }^{N_p}
\textrm{dim} \ V_{p \alpha_p} = \sum_{p=0}^n (N_p -p +1) \
\textrm{dim} \ V_p \,.\,
\end{eqnarray}
The structure constants of ${\check \G}_S^\prime$ are given by
\begin{eqnarray}
\label{S-C}
 C_{i_p,\beta_p \; j_q,\gamma_q}^{k_s,\alpha_s}&=&
\delta_{\beta_p\gamma_q}^{\alpha_s}c_{i_p j_q}^{k_s},\quad
p,q,s=0,1,\ldots, n \;, \quad \alpha_p,\beta_p,\gamma_p=p,\ldots,N_p
\\&=&\left\{
\begin{array}{lll} 0, &
\mathrm{if} \ \beta_p + \gamma_q \neq \alpha_s  \\
c_{i_p j_q}^{k_s}, & \mathrm{if} \ \beta_p + \gamma_q = \alpha_s
\end{array} \right. \quad
\begin{array}{l}
p,q,s=0,1,\ldots, n \\
i_{p,q,s}=1,2,\ldots, \textrm{dim} \, V_{p,q,s} \\
\alpha_p,\beta_p,\gamma_p=p,p+1, \ldots, N_p \quad ,
\end{array}\label{scWWn}
\end{eqnarray}
where the part $\delta_{\beta_p\gamma_q}^{\alpha_s}$ of the
structure constants (see (\ref{scdirsum1})), in which and
$\beta_p$, $\gamma_q$ ..., indicate the elements of the subsets
${\check S}_p$, ${\check S}_q$ ... above,  is obtained from
(\ref{SN1mult}). We see that the dimensions in eqs.
(\ref{dimcheck}) and (\ref{eq:dim}), and the structure constants
in eqs. (\ref{scWWn}) and (\ref{eq:Cn}), coincide. Thus, if the
integers $N_p$ are restricted as in (\ref{conWW}), the above
algebra ${\check \G}_S^\prime$ is just the expansion ${\cal
G}(N_0,\ldots,N_n)$ \cite{AIPV} of Theorem 1.

We refer to \cite{IzRoSal06} for further details on $S$-expanded
algebras.
%
%
% %${\check \G}_S^\prime \subset \G_S^\prime$ by
%\begin{equation}
%{\check \G}_S^\prime \; = \; \oplus_{p} \left( \oplus_{\alpha_p
%\in {\check S}_p} V_{p \alpha_p} \right), \qquad p=0,\ldots,n  ,
%\end{equation}
%({\it c.f.} (\ref{vsgps}); ${\check \G}_S^\prime$ is not a
%subalgebra of ${\G}_S^\prime$).  ${\check \G}_S^\prime$ turns out
%to be a Lie algebra with structure constants
%\begin{eqnarray}
%\label{S-C}
% C_{p,\beta_p \; q,\gamma_q}^{s,\alpha_s}&=&
%\delta_{\beta_p\gamma_q}^{\alpha_s}c_{pq}^{s},\quad
%p,q,s=0,1,\ldots, n \;, \quad
%\alpha_p,\beta_p,\gamma_p=p,\ldots,N_p
%\\&=&\left\{
%\begin{array}{lll} 0, &
%\mathrm{if} \ \beta_p + \gamma_q \neq \alpha_s  \\
%c_{pq}^{s}, & \mathrm{if} \ \beta_p + \gamma_q = \alpha_s
%\end{array} \right. \quad
%\begin{array}{l}
%p,q,s=0,1,\ldots, n \\
%\alpha_p,\beta_p,\gamma_p=p,\ldots,N_p \qquad ,
%\end{array}\label{scWWn}
%\end{eqnarray}
%where the part $\delta_{\beta_p\gamma_q}^{\alpha_s}$ of the
%structure constants (see (\ref{scdirsum1})), in which and
%$\beta_p$, $\gamma_q$ ..., indicate the elements of the subsets
%${\check S}_p$, ${\check S}_q$ ... above,  is obtained from
%(\ref{SN1mult}). We now see that (\ref{scWWn}) gives the structure
%constants of (\ref{eq:Cn}). Thus, the algebra ${\check
%\G}_S^\prime$ is just the expansion ${\cal G}(N_0,\ldots,N_n)$
%\cite{AIPV} of Theorem 1. We refer to \cite{IzRoSal06} for other
%S-expanded algebras.
%
%

\section{The gauge structure of $D=11$ supergravity} \label{sugra}

As a recent physical application of the expansion method, we now
comment briefly on the underlying gauge structure of
eleven-dimensional supergravity \cite{D'A+F82,BAIPV04}. See
\cite{HS,AIPV,
AIPV04,Hatsuda:2003ry,Hatsuda:2004vi,Sakaguchi:2006pg, Izqunpub,
Edelstein:2006se, Edelstein:2006cd} for other possible applications
of the expansion method.

We are interested here in the underlying gauge symmetry of $D=11$
Cremmer-Julia-Scherk (CJS) supergravity \cite{CJS} as a way of
understanding the symmetry structure of M-theory, the low energy
limit of which is $D$=11 supergravity. The problem of its hidden
or underlying gauge geometry was raised already in the CJS
pioneering paper \cite{CJS}, where the possible relevance of
$OSp(1|32)$ was suggested. It was specially considered by D'Auria
and Fr\'e \cite{D'A+F82}, who looked at the problem as a search
for a composite structure of its three-form field $A_3(x)$.
Indeed, while two of the $D=11$ supergravity fields (the graviton
$e^a=dx^\mu e_\mu^a(x)$ and the gravitino $\psi^\alpha= dx^\mu
\psi^\alpha_\mu(x)$) are given by {\it one}-form spacetime fields
and thus can be considered, together with the spin connection
($\omega^{ab}=dx^\mu\omega_\mu^{ab}(x)$), as gauge fields for the
standard superPoincar\'e group, the additional
$A_{\mu_1\mu_2\mu_3}(x)$ abelian gauge field in $D=11$ CJS
supergravity is not associated with any superPoincar\'e algebra MC
{\it one}-form or  generator since it rather corresponds to a {\it
three}-form $A_3$. However, one may ask whether it is possible to
introduce a set of additional one-form fields associated to the LI
MC forms of a larger superalgebra such that these fields, together
with $e^a$ and $\psi^\alpha$, can be used to express $A_3$ in
terms of one-forms. If so, the `old' $e^a,\psi^\alpha$ and the
`new' one-form fields may be considered as gauge fields of a
larger supersymmetry group, with $A_3$ expressed in terms of them.
This is what is meant by the underlying gauge group structure of
CJS supergravity: it is hidden when the standard $D=11$
supergravity multiplet is considered, and manifest when the
three-form field $A_3$ becomes a {\it composite} of one-form
fields associated with the MC forms of the larger superalgebra, in
which case all CJS supergravity fields can be treated as one-form
gauge fields. It is then seen that the solution of this problem is
equivalent to trivializing a standard $D=11$ supersymmetry algebra
$\mathfrak{E}^{(11|32)}$ cohomology four-cocycle $\omega_4$
(structurally equivalent to the four-form $dA_3$) on a {\it
larger} algebra $\tilde{\mathfrak{E}}$ corresponding to a larger
superspace group ${\tilde{\Sigma}}$.

It turns out \cite{BAIPV04} that there is a whole one-parameter
family of enlarged supersymmetry algebras
$\tilde{\mathfrak{E}}(s), s\not=0$ that trivialize the
$\mathfrak{E}^{(11|32)}$ four-cocycle $\omega_4$ ($\sim dA_3$)
(see \cite{BAIPV04} for the meaning of $\tilde{\mathfrak{E}}(s)$
and its associated family of enlarged superspace groups
${\tilde{\Sigma}}(s)$). Hence (and adding the $D=11$ Lorentz
group, $SO(1,10)$), this means that the underlying gauge
supergroup of $D=11$ supergravity has a semidirect structure and
can be described by any representative of a {\it one-parametric
family of supergroups, ${\tilde{\Sigma}(s)
\times\!\!\!\!\!\!\supset SO(1,10)}$ for $s\not=0$}. These may be
seen as deformations of ${\tilde{\Sigma}(0)
\times\!\!\!\!\!\!\supset SO(1,10)} \subset {\tilde{\Sigma}(0)
\times\!\!\!\!\!\!\supset Sp(32)} $, where $\tilde{\Sigma}(0)$ is
a certain enlarged superspace group \cite{BAIPV04}. Thus our
conclusion is that the underlying gauge group structure of $D=11$
supergravity is determined by a one-parametric nontrivial
deformation of ${\tilde{\Sigma}(0)\times\!\!\!\!\!\!\supset
SO(1,10)}$ $\subset$ ${\tilde{\Sigma}(0)\times\!\!\!\!\!\!\supset
Sp(32)}$ (two specific cases of the $\tilde{\mathfrak{E}}(s)$
family, $\tilde{\mathfrak{E}}(3/2)$ and
$\tilde{\mathfrak{E}}(-1)$, were already found in \cite{D'A+F82}).
The singularity of $\tilde{\mathfrak{E}}(0)$ looks reasonable; the
corresponding $\tilde{\Sigma}(0)$ enlarged superspace group is
special because the Lorentz $SO(1,10)$ automorphism group of
$\tilde{\Sigma}(s)$ ($s\not=0$) is enhanced to $Sp(32)$ for
$\tilde{\Sigma}(0)$. The appearance of $\tilde{\Sigma}(0)$ allows
us to clarify the connection of the underlying gauge supergroups
with $OSp(1|32)$ above mentioned. It is found \cite{BAIPV04} that
${\tilde{\Sigma}(0) \times\!\!\!\!\!\!\supset SO(1,10)}$ is an
expansion of $OSp(1|32)$; specifically, ${\tilde{\Sigma}(0)
\times\!\!\!\!\!\!\supset SO(1,10)} \approx OSp(1|32)(2,3,2)$. It
may also be shown that ${\tilde{\Sigma}(0)
\times\!\!\!\!\!\!\supset Sp(32)}$ is the expansion of
$OSp(1|32)(2,3)$.

The enlarged supersymmetry algebras $\tilde{\mathfrak{E}}(s)$ are
central extensions of the M-algebra (of generators $Q_\alpha,P_a,
Z_{ab}, Z_{a_1\ldots a_5}$) by an additional fermionic generator
$Q^\prime_\alpha$. Trivializing the $\mathfrak{E}^{(11|32)}$ Lie
superalgebra cohomology four-cocycle $\omega_4$ on the enlarged
supersymmetry algebra $\tilde{\mathfrak{E}}(s)$, so that
$\omega_4$ is the exterior derivative of an invariant form,
$\omega_4 = d\tilde{\omega}_3$, is tantamount to finding a
composite structure for the three-form field $A_3$ of CJS
supergravity in terms of one-form gauge fields, $A_3=A_3(e^a \, ,
\, \psi^\alpha \; ; \; B^{a_1a_2}, B^{a_1 \ldots a_5} ,
\eta^\alpha \,)$ associated to the MC forms of
$\tilde{\mathfrak{E}}(s)$. The compositeness of $A_3$ is given by
the same equation that provides the $\tilde{\omega}_3$
trivialization $\omega_4=d\tilde{\omega}_3$ of the $\omega_4$
cocycle (where {\it now} $\tilde{\omega}_3$ is
$\tilde{\Sigma}(s)$-invariant; this is why $\omega_4$ becomes a
{\it trivial} cocycle for $\tilde{\mathfrak{E}}(s)$, $s\not=0$;
see {\it e.g.} \cite{CUP}). In the composite $A_3$ expression, the
$\tilde{\mathfrak{E}}(s)$ MC forms are replaced by `soft'
one-forms -spacetime one-form fields- obeying a free differential
algebra with curvatures.

  The presence of the additional one-form gauge fields associated with
the new generators in $\tilde{\mathfrak{E}}(s)$ might be expected.
The field $B^{a_1\ldots a_5}$, associated to the $Z_{a_1\ldots
a_5}$ M-algebra generator, is needed \cite{BPS04} for a coupling
to BPS preons \cite{BAIL01}, the hypothetical basic constituents
of M-theory. In a more conventional perspective, one can notice
that the generators $Z_{a_1 a_2}$ and $Z_{a_1\ldots a_5}$ can be
treated as topological charges \cite{AGIT89} of the M2 and M5
superbranes (see also \cite{ST97}). In the standard CJS
supergravity the M2-brane solution carries a charge of the
three-form gauge field $A_3$ and thus there should have a relation
with the charge $Z_{a_1 a_2}$ and its gauge field $B^{a_1 a_2}$.
The analysis of the r\^ole of the fermionic central charge
$Q^\prime_\alpha$ and its gauge field $\eta^\alpha$ in this
perspective requires more care, although such a fermionic
`central' charge is also present in the Green algebra \cite{Gr89}
(see also \cite{BeSe95,Se97,HaSa00, CAIPB00} and references
therein).

\vspace{0.5cm}

Some comments are now in order.

\begin{itemize}

\item The supergroup manifolds $\tilde{\Sigma}(s)$ are
\emph{enlarged} superspaces. The fact that all the {\it spacetime}
fields appearing in the above description of CJS supergravity may
be associated to the various coordinates of $\tilde{\Sigma}(s)$ is
suggestive of an {\it enlarged superspace variables/spacetime
fields correspondence} principle for $D=11$ CJS supergravity.

\item This is not the only case where such a situation appears.
It may be seen \cite{CAIPB00} that one may introduce an {\it
enlarged superspace variables/worldvolume fields correspondence}
principle for superbranes, by which one associates all {\it
worldvolume} fields, including the Born-Infeld (BI) ones
\cite{CAIPB00, Sa99} in the various D-brane actions, to fields
corresponding to forms defined on suitably enlarged superspaces
$\tilde{\Sigma}$ (the actual worldvolume fields are the pull-backs
of these forms to the worldvolume of the extended supersymmetric
object). The worldvolume BI fields, as the spacetime $A_3$ field
of CJS supergravity above, become composite fields. Moreover, a
Chevalley-Eilenberg Lie algebra cohomology analysis
\cite{AT89,CAIPB00,AI04} of the Wess-Zumino terms of many
different superbrane actions determines the possible ones and how
the ordinary supersymmetry algebra has to be extended (see also
\cite{Sa99,Ha98}). This again suggests an enlarged superspace
variables/worldvolume fields correspondence.

\item Thus, could there be an {\it enlarged superspace
variables/fields correspondence principle in M-theory}?

\end{itemize}

To conclude, we would like to mention that the expansion method
can also be applied \cite{AIPV} to free differential algebras
(FDAs) \cite{Su77,D'A+F82,Ni83,Cd'AF91}, structures that prove
useful to discuss the dynamics of supergravity theories. In
particular, it can be applied to the {\it gauge} FDAs obtained by
`softening' the MC forms, and therefore to obtain  Chern-Simons
type actions, from those for the unexpanded algebras
\cite{AIPV,Izqunpub,Edelstein:2006se} (see \cite{Edelstein:2006cd}
for a review of Chern-Simons actions in the supergravity context).
The $S$-expansions \cite{IzRoSal06} briefly reviewed at the end of
Sec.~\ref{exp} can also be applied to the construction of
Chern-Simons lagrangians \cite{Izaurieta:2006aj}. The reduction of
the $D=11$ supergravity FDA has been very recently analyzed
\cite{Vaula:2006hv} in terms of the Sezgin algebra \cite{Se97} and
the $E_{11}$ Kac-Moody algebra.

 \vspace{0.5cm}
{\bf Acknowledgments}. The authors wish to thank I.~Bandos  for
useful discussions and E. Weimar-Woods for correspondence. This
work has been partially supported by research grants from the
Spanish Ministerio de Educaci\'on y Ciencia (FIS2005-02761,
FIS2005-03989)and EU FEDER funds, the Generalitat Valenciana, the
Junta de Castilla y Le\'on (VA013C05) and the EU `Forces Universe'
network (MRTN-CT-2004-005104). M.P. and O.V. would like to thank,
respectively, the Spanish Ministerio de Educaci\'on y Ciencia and
the Generalitat Valenciana for the FPU and FPI research
fellowships they held whilst this work was being carried out.

\end{document}